# Spontaneous and Stimulated Emissions of Quantum Free-Electron Wavepackets - QED Analysis

Yiming Pan, Avraham Gover

*Department of Electrical Engineering Physical Electronics,*

*Tel Aviv University, Ramat Aviv 69978, ISRAEL*

## Abstract

Do the wavepacket-size of free-electron wavefunction and its history have physical effect in its interaction with light? Here we answer this problem by analyzing a QED model, considering both spontaneous and stimulated emission of quantized radiation field. For coherent radiation (Glauber state), we confirm that stimulated emission/absorption of photons has a dependence on wavepacket size that decays when it exceeds the interacting radiation wavelength, consistently and complementarily with Schrodinger equation analysis of wavepacket acceleration in classical electromagnetic field. Furthermore, the stimulated emission of modulated electron wavepacket with coherently-bunched profiles has characteristic harmonic emission spectrum that is also wavepacket size dependent but beyond the frequency cut-off. In either case, there is no wavepacket dependent emission of Fock state radiation, and particularly the vacuum state spontaneous emission is wavepacket-independent. The transition of radiation emission from the classical point-particle limit to the quantum electron wavefunction limit is demonstrated in electron wavepacket representation. It indicates a way for measuring the wavepacket size of single electron wavefunction, and suggests a new direction for exploring light-matter interaction fundamentally.

Accelerated free electrons emit electromagnetic radiation when subjected to an external force (e.g. synchrotron radiation [1], Undulator radiation [2], Compton scattering [3]). Radiation can also be emitted by currents that are induced by free electrons in polarizable structures and materials, such as in Cherenkov radiation [4], transition radiation [5], Smith-Purcell radiation [6]. Some of these schemes were demonstrated to operate as coherent stimulated radiative emission sources, such as Free Electron Lasers (FEL) [7-9], as well as accelerating (stimulated absorption) devices, such as Dielectric Laser Accelerator (DLA) and Inverse Smith-Purcell effect [10-12].

Most of the free electron radiation schemes of emission or acceleration operate in the classical theoretical regime of electrodynamics, where the electrons can be considered point-particles and the radiation field is described by Maxwell equations (no field quantization). However, a variety of free electron radiation schemes [15,16], and particularly FEL [e.g. Refs: 17,18,19] have been analyzed in the framework of a quantum model in which the electron is described in the inherently quantum limit, given as a plane-wave quantum wavefunction – the opposite limit of the point-particle classical presentation. Quantum description of the electron wavefunction is also used in another recently developed research field of electron interaction with radiation: Photo-Induced Near-Field Electron Microscopy (PINEM) [20,21]. In this scheme, a single electron quantum wavefunction interacts with the near-field of a nanometric structure illuminated by a coherent laser beam. Of special relevance for the present discussion is a recent PINEM-kind experiment of Feist et al [22], in which it was demonstrated that optical frequency modulation of the energy and density expectation values of a single electron wavepacket are possible in this method.

The extremely different presentations of the radiative interaction of free electron with radiation in the classical and quantum limits raise interest in the theoretical understanding of the transition of the electron-radiation interaction process from the quantum to the classical limit. This is also related to deeper understanding of fundamental physics questions, such as the particle-wave duality nature of the electron [23].

In the classical description, the point-particle dynamics is governed by the Lorentz force equation and its radiation – by Maxwell equations. The radiation field emitted by a single electron spontaneously in free space, and the stimulated emission/absorption of an incident radiation field depend on the well-defined entrance time $t_{0e}$ of the electron to the interaction region (namely, its entrance phase relative to the radiation wave $\varphi(t_{0e}) = \omega t_{0e}$). However, "Classical spontaneous emission" of free electrons (e.g. "Undulator radiation" [2], "Cerenkov radiation" etc.) is only described in the context of an ensemble (pulse) of multi-particles injected randomly into the interaction region, and therefore, after averaging over the ensemble there is no dependence left on the phase of individual electrons. Also in superradiant coherent spontaneous emission by an ensemble of bunched electrons [14] and in classical description of stimulated radiative emission schemes, such as FEL [8], the phase dependence of the individual electrons disappears after averaging. On the other hand, in the quantum description of spontaneous and stimulated radiation field by a free electron there is no phase dependence already at the level of a single electron, because the electron is described by an infinitely extended plane-wave [15]. The spontaneous emission is described as a consequence of "zero-field vibration" in a quantized field QED model. The stimulated radiation emission/absorption is explained in terms of multi-photon emission/absorption processes in which the electron makes transition to lower/higher energy states of its continuous energy dispersion curve. Only because this energy dispersion curve is nonlinear, the two processes are not degenerate and do not overlap (in the quantum interaction limit) or only partly overlap (in the classical interaction limit) and net stimulated interaction gain/loss of the radiation wave is possible [15].

The way to settle these two diverse points of view of the electron – radiation interaction, and understand their classical-to-quantum transition, is to describe the free electron as a quantum wavepacket which would tend to resemble a plane wave when the wavepacket is long relative to the radiation wavelength, and a point particle in the opposite limit. This model has been presented by us in separate semi-classical quantum formulations in which the radiation field is a mode expansion classical field solution of

Maxwell equations. In the electron dynamics problem of stimulated emission, the electron wavefunction is the solution of Schrodinger equation [27], and in the radiation problem, the current source is related to the expectation value of the quantum wavepacket probability density $|\psi(r,t)|^2$ [28]. Both formulations are consistent with each other in the case of stimulated emission, and are consistent with the point particle and quantum electron limits. They show wavepacket phase and size dependent transition of stimulated emission/absorption (deceleration/acceleration) in the short wavelength limit to null emission and acceleration in the long quantum electron wavepacket limit. The spontaneous emission of the wavepacket shows similar transition and radiation cutoff in the long wavepacket size limit.

In this article, we analyze the spontaneous and stimulated emission problem of a quantum electron wavepacket and a modulated wavepacket in a QED model. The more general QED model is consistent with the semi-classical analyses in the case of stimulated interaction with a coherent (Glouber state) radiation field, but as expected, has different predictions in the case of spontaneous emission. Since recent experimental progress makes it feasible to generate, accelerate, control the shape and size and modulate single electron wavepackets, we assert, based on the presented theory, that radiative interaction experiments in the transition range between classical to quantum electron wavepacket limits are a viable way for measuring the dimension and structure of electron quantum wavepacket. It can help to resolve the difference in description of spontaneous emission in the classical and quantum formulations, and offers a new way for studying fundamental aspects of radiation - matter interaction in the quantum limits.

## Modeling and Methods

Our QED analysis is based on first order perturbation solution of the relativistically modified Schrodinger equation [15, 27] for a free electron wavefunction and a quantized radiation field. The unperturbed Hamiltonian is similar to the one used in conventional quantum analysis of free electron interaction [15], but as in [27], the equation is solved here with initial conditions of a finite size electron quantum wavepacket instead of a plane wave. The interaction Hamiltonian is taken to be:

$$H_I(t) = -\frac{e\left(\hat{\mathbf{A}} \cdot (-i\hbar\nabla) + (-i\hbar\nabla) \cdot \hat{\mathbf{A}}\right)}{2\gamma_0 m} \qquad (1)$$

For the case of our concern,

$$\hat{\mathbf{A}} = -\frac{1}{2i\omega}\left(\hat{\mathbf{E}}(\mathbf{r})e^{-i\omega t} - \hat{\mathbf{E}}^\dagger(\mathbf{r})e^{i\omega t}\right) \qquad (2)$$

where $\hat{\mathbf{A}}, \hat{\mathbf{E}}$ are field operators. In our one-dimensional analysis, we assume that the interaction of an electron of velocity $v_0$ with a single radiation mode q takes place through an axial slow-wave field component of the mode: $\hat{\mathbf{E}}(\mathbf{r}) = \sum_\nu \mathcal{E}_{qz0} e^{iq_z z - i\phi_0} \hat{a}_{q\nu} \hat{\mathbf{e}}_z$ that is nearly synchronous with the electron: $v_0 = \omega/q_z$ and $\hat{a}_q (\hat{a}_q^\dagger)$ is the annihilation (creation) operator of photon number state $\nu$ in this mode. This slow wave component may the axial component of the field of one of the space-harmonics of a classical Floquet-mode radiation wave incident on a grating in a Smith-Purcell structure [27,28]: $\tilde{\mathcal{E}}_q(\mathbf{r}) = \sum_m \tilde{\mathcal{E}}_{qm}(\mathbf{r}_\perp) e^{iq_{zm}z}$, where $q_{zm} = q_{z0} + m2\pi/\lambda_G$. In this case, and in other examples, such as Cerenkov radiation [4,16] or interaction with the evanescent field of a wave guided in a dielectric waveguide [9], the slow traveling wave component can be related to the total normalization power of the radiation mode $\mathcal{P}_q$ through a "Pierce impedance" parameter [32]

$$K_q = \mathcal{E}_{qz0}^2 / 2q_{z0}^2 \mathcal{P}_q \qquad (2A)$$

In the radiation quantization, we quantize the energy carried by the radiation mode during a time period $t_r = L/v_0$ along an interaction length L:

$$W_{q\nu} = \mathcal{P}_q t_r \nu = \hbar\omega\nu \qquad (2B)$$

Therefore, as we solve the Schrodinger equation for the interaction Hamiltonian (1) with the field of the slow-wave component, the relations (2A, 2B) would produce the photon emission increment of the entire radiation mode.

Following the standard quantum electrodynamics theory, we expand the initial wave function in terms of the quantum numbers p of the electron state and the Fock photon occupation state of mode q, which is given by:

$$|i\rangle = \sum_{p,v} c^{(i)}_{p,v}(t) |p,v\rangle \qquad (3)$$

During light-matter interaction, the electron and radiation field exchange energy and momentum, and evolve to a final combined state:

$$|f\rangle = \sum_{p,v} c^{(f)}_{p,v}(t) |p,v\rangle \qquad (4)$$

with combined state coefficients $c^{(f)}_{p,v}(t)$. For the case of an electron wavepacket in our one-dimensional model, the initial wavefunction is given by:

$$|i\rangle = \int \frac{dp}{\sqrt{2\pi\hbar}} \sum_v c^{(0)}_{p,v} e^{-iE_p t/\hbar} |p,v\rangle \qquad (5)$$

where the energy dispersion relation of relativistic free-electrons is $E_p = c\sqrt{m^2 c^2 + p^2}$. First order time-dependent perturbation theory of Schrodinger equation [29] results in:

$$i\hbar \dot{c}^{(1)}_{p',v'} = \int \frac{dp}{\sqrt{2\pi\hbar}} \sum_v c^{(0)}_{p,v} \langle p',v'|H_I(t)|p,v\rangle e^{-i(E_p - E_{p'})t/\hbar} \qquad (6)$$

Integrating (eq.5) in time t from 0 to infinity, the emission and absorption process terms of the first order perturbation coefficient $c^{(1)}_{p',v'} = c^{(1)(e)}_{p',v'} + c^{(1)(a)}_{p',v'}$ are respectively:

$$c^{(1)(e,a)}_{p',v'} = \frac{\pi}{2i\hbar} \int \frac{dp}{\sqrt{2\pi\hbar}} \sum_v c^{(0)}_{p,v} \langle p',v'|H^{(e,a)}_I(0)|p,v\rangle \delta\left(\frac{E_p - E_{p'} \mp \hbar\omega}{2\hbar}\right) \qquad (7)$$

where $H^{(e,a)}_I$ correspond respectively to the second and first terms in the interaction Hamiltonian (1).

The momentum quantum recoil of the electron is found from substitution in (7) of the energy dispersion relation expanded to second order:

$$E_p = c\sqrt{m^2c^2 + p^2} \approx \varepsilon_0 + v_0(p - p_0) + \frac{(p-p_0)^2}{2m^*}.$$ Determined by the delta functions, it is:

$$p_{rec}^{(e,a)} = p'^{(e,a)} - p_0 = p_{rec}^{(0)}(1 \pm \delta) \tag{8}$$

where $p_{rec}^{(0)} = \hbar\omega/v_0$, $\delta = \hbar\omega/2m^*v_0^2$, $m^* = \gamma_0^3 m$. Then the first order perturbation coefficient is given by:

$$c_{p',v'}^{(1)(e,a)} = \frac{\pi}{iv_0} \sum_v c_{p' \pm p_{rec}^{(e,a)}, v}^{(0)} \langle p', v' | H_I^{(e,a)}(0) | p' \pm p_{rec}^{(e,a)}, v \rangle \tag{9}$$

and the general matrix element is given explicitly by:

$$\langle p', v' | H_I^{(e,a)}(0) | p, v \rangle = \begin{cases} + \frac{e\hbar\mathcal{E}_{qz0}}{2\gamma_0 m\omega} \langle v' | \hat{a}_q^\dagger | v \rangle \cdot \int \frac{dz}{2\pi\hbar} e^{-i(q_z z - \phi_0)} e^{-ip'z/\hbar} \left( \frac{\partial}{\partial z} - iq_z/2 \right) e^{ipz/\hbar} \\ - \frac{e\hbar\mathcal{E}_{qz0}}{2\gamma_0 m\omega} \langle v' | \hat{a}_q | v \rangle \cdot \int \frac{dz}{2\pi\hbar} e^{i(q_z z - \phi_0)} e^{-ip'z/\hbar} \left( \frac{\partial}{\partial z} + iq_z/2 \right) e^{ipz/\hbar} \end{cases} \tag{10}$$

Therefore,

$$c_{p',v'}^{(1)(e,a)} = \begin{cases} + \left( \frac{p' + p_{rec}^{(e)} - \hbar q_z/2}{p_0} \right) \Upsilon \sqrt{v'} c_{p'+p_{rec}^{(e)}, v'-1}^{(0)} \text{sinc}\left(\bar{\theta}^{(e)}/2\right) e^{i(\bar{\theta}^{(e)}/2 + \phi_0)} \\ - \left( \frac{p' - p_{rec}^{(a)} + \hbar q_z/2}{p_0} \right) \Upsilon \sqrt{v'+1} c_{p'-p_{rec}^{(a)}, v'+1}^{(0)} \text{sinc}\left(\bar{\theta}^{(a)}/2\right) e^{-i(\bar{\theta}^{(a)}/2 + \phi_0)} \end{cases} \tag{11}$$

with $\Upsilon = \frac{e\mathcal{E}_{qz0}L}{4\hbar\omega}$, $p_0 = \gamma_0 m v_0$, and

$$\bar{\theta}^{(e,a)} = \frac{(p_{rec}^{(e,a)} \pm \hbar q_z)L}{\hbar} = \left( \frac{\omega}{v_0} - q_z \right)L \pm \left( \frac{\omega}{v_0} \right)L\delta = \bar{\theta} \pm \frac{\varepsilon}{2} \tag{12}$$

where $\bar{\theta} = \left( \frac{\omega}{v_0} - q_z \right)L$ is the classical interaction 'detuning parameter', $\varepsilon = \delta\left( \frac{\omega}{v_0} \right)L \ll 1$ is the interaction quantum recoil parameter [15].

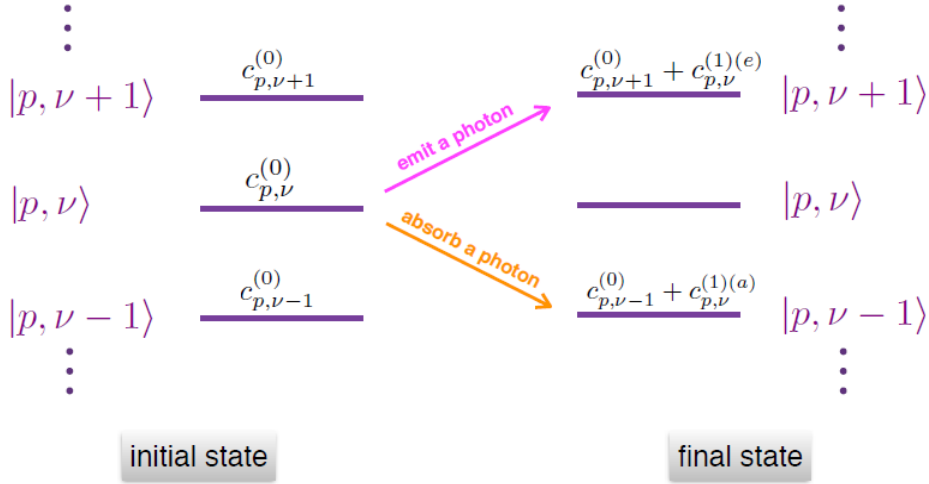

Fig.1: *The schematic diagram shows the light-matter scattering processes of emitting and absorbing a photon from an initial electron-photon distribution $\left|c_{p,\nu}^{(i)}\right|^2$ to a final re-distribution $\left|c_{p,\nu}^{(f)}\right|^2$.*

As shown in Fig.1, the schematic diagram shows the light-matter scattering processes of emitting and absorbing a photon. Explicitly, for emitting a photon, the final coefficient of state $|p,\nu+1\rangle$ is given by $c_{p,\nu+1}^{(0)} + c_{p,\nu+1}^{(1)(e)}$. This represents a reciprocal electron momentum and energy conserving process through emission of a photon and momentum back recoil: $|p+p_{rec}^{(e)},\nu\rangle \Rightarrow |p,\nu+1\rangle$ [28]. On the other hand, for absorbing a photon, the final coefficient of state $|p,\nu-1\rangle$ is given by $c_{p,\nu-1}^{(0)} + c_{p,\nu-1}^{(1)(a)}$, which corresponds to the process of absorbing a photon and electron momentum forward recoil: $|p-p_{rec}^{(a)},\nu\rangle \Rightarrow |p,\nu-1\rangle$. Finally, the net photon emission/absorption is derived from

$$\Delta\nu_q = \sum_{p,\nu}\left(\left|c_{p,\nu+1}^{(0)} + c_{p,\nu+1}^{(1)(e)}\right|^2 - \left|c_{p,\nu-1}^{(0)} + c_{p,\nu-1}^{(1)(a)}\right|^2\right) \tag{13}$$

This can be expressed as the sum of two terms $\Delta\nu_q = \Delta\nu_q^{(1)} + \Delta\nu_q^{(2)}$, i.e.,

$$\Delta v_q^{(1)} = 2\sum_{p,v} \Re\left\{\left(c_{p,v+1}^{(0)*} \cdot c_{p,v+1}^{(1)(e)}\right) - \left(c_{p,v-1}^{(0)*} \cdot c_{p,v-1}^{(1)(a)}\right)\right\}$$

$$\Delta v_q^{(2)} = \sum_{p,v}\left(\left|c_{p,v+1}^{(1)(e)}\right|^2 - \left|c_{p,v-1}^{(1)(a)}\right|^2\right)$$

(14)

Note that from now on we replace the index $p' \to p$ for the final momentum distributions (eq.11). The second term $\Delta v_q^{(2)}$ is the same as the expression that has been derived in previous QED formulations of photon emission by free electrons in the infinite quantum wavepacket limit using conventional Fermi's golden rule analysis [15]. The first term $\Delta v_q^{(1)}$ is the contribution from the interference between the initial and scattered states that depends on the features of the initial wavefunction distribution. It has not been considered in previous analysis, and is an innovation of the present formulation.

## Results

In the present analysis, we consider the case where the electron wavefunction and the radiation field are initially disentangled: $c_{p,v}^{(0)} = c_p^{(0)} c_v^{(0)}$. Substitution of (11) in (14) results in then:

$$\Delta v_q^{(1)} = \sum_{p,v}\left(\sqrt{v+1}\,\rho_{p,v}^{(1)(e)} + \sqrt{v}\,\rho_{p,v}^{(1)(a)}\right)$$

$$\Delta v_q^{(2)} = \sum_{p,v}\left((v+1)\rho_{p,v}^{(2)(e)} - v\rho_{p,v}^{(2)(a)}\right)$$

(15)

where

$$\rho_{p,v}^{(1)(e,a)} = 2\Upsilon\left(\frac{p \pm (p_{rec}^{(e,a)} \mp \hbar q_z/2)}{p_0}\right)\mathrm{sinc}\left(\bar{\theta}^{(e,a)}/2\right)\cos\left(\bar{\theta}^{(e,a)}/2 + \phi_0\right)\left|c_p^{(0)*}c_{p \pm p_{rec}^{(e,a)}}^{(0)} c_{v \pm 1}^{(0)*} c_v^{(0)}\right|$$

$$\rho_{p,v}^{(2)(e,a)} = \Upsilon^2\left(\frac{p \pm (p_{rec}^{(e,a)} \mp \hbar q_z/2)}{p_0}\right)^2 \mathrm{sinc}^2\left(\bar{\theta}^{(e,a)}/2\right)\left|c_{p \pm p_{rec}^{(e,a)}}^{(0)}\right|^2 \left|c_v^{(0)}\right|^2$$

(16)

We are set now to examine various cases of interest: (a) spontaneous emission; (b) stimulated emission with quantum light, and particularly with a single Fock state - $c_v^{(0)} = \delta_{v,v_0}$, and (c) stimulated emission from a coherent Glauber state.

Spontaneous Emission

In this case

$$c_v^{(0)} = \delta_{v,0} \tag{17}$$

and we get from the second order perturbation terms of equations (16,17) that the only nonzero quantum transition term is single photon emission from the vacuum state (see Fig. 2):

$$\Delta v_{q,\mathrm{SP}} = \Delta v_q^{(2)}\big|_{v=0} = \sum_p \left| c_{p,1}^{(1)(e)} \right|^2 = \Upsilon^2 \mathrm{sinc}^2\left(\bar{\theta}^{(e)}/2\right) \tag{18}$$

where we approximated $p_{rec}^{(e)}, \hbar q_z / p_0 \ll 1$.

Remarkably, in this case, (15, 17) produce null result $\Delta v_q^{(1)} = 0$ for the spontaneous emission contribution of the first order perturbation term. Eq. 18 gives the only source of spontaneous photons emission, and therefore, there is no wavepacket size or shape dependence of spontaneous emission!

The spontaneous emission rate can be found by dividing (18) by the interaction transit time $L/v_0$

$$\left(\frac{dv_q}{dt}\right)_{\mathrm{SP}} = \left(\frac{v_0}{L}\right) \Upsilon^2 \mathrm{sinc}^2\left(\bar{\theta}^{(e)}/2\right) \tag{19}$$

Consistently with the general well-established expression for spontaneous emission by an infinite (plane wave) quantum electron wavefunction [15]. Note, however, that in this case of slow wave interaction, $\Upsilon = e\mathcal{E}_{qz0}L/4\hbar\omega$ is given in terms of the axial slow field component of the normalized mode $\mathcal{E}_{qz0}$. Spontaneous emission into the entire mode $\Delta v_{q,\mathrm{SP}}$ (18) or the spontaneous energy emission $\Delta W_{q,\mathrm{SP}} = \hbar\omega\Delta v_{q,\mathrm{SP}}$ can be explicitly calculated from the total mode energy normalization relation (2A, 2B).

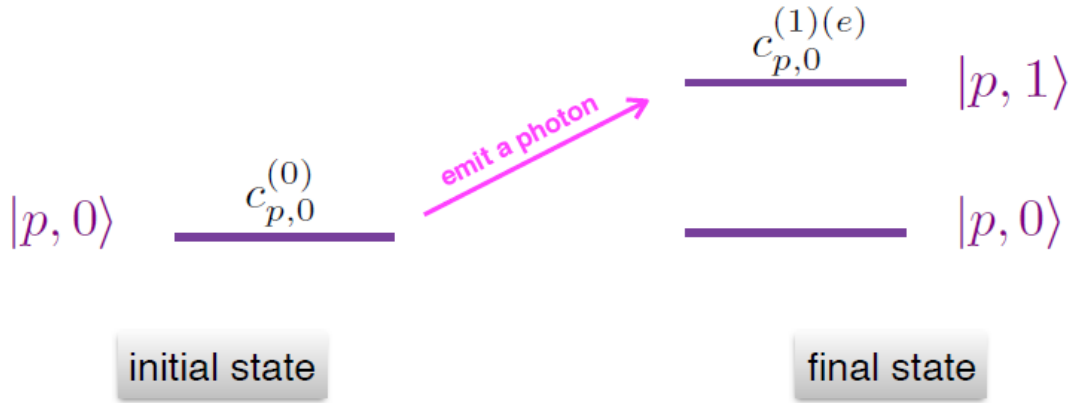

Fig.2: *The spontaneous emission occurs when the electromagnetic field at vacuum state in the absence of any light sources.*

Stimulated Emission – Fock photon state

In this case

$$c_v^{(0)} = \delta_{v,v_0} \qquad (20)$$

Inspecting (15,16), it is straightforward to see that similarly to the case of spontaneous emission (which is simply the case $v_0 = 0$), there is no Fock-state stimulated emission due to the first order terms; namely, $\Delta v_q^{(1)}\big|_{v_0} = 0$, because their substitution result in null terms: $\sqrt{v_0+1}\left(c_{v_0+1}^{(0)*}c_{v0}^{(0)}\right) = \sqrt{v_0}\left(c_{v_0-1}^{(0)*}c_{v_0}^{(0)}\right) = 0$. There is therefore no stimulated radiative interaction with a Fock-state radiation wave. This is hardly surprising, since a Fock-state wave has no phase.

The second order terms in (15, 16) do produce wavepacket independent stimulated emission. Again, with the approximation $p_{rec}^{(e)}, \hbar q_z / p_0 \ll 1$, and the limit of an infinite (plane wave) electron wavepacket $c_p^{(0)} = \delta(p - p_0)$ the momentum integration in (15b) results in:

$$\Delta v_q^{(2)} = \Upsilon^2 \left((v_0 + 1)\operatorname{sinc}^2\left(\overline{\theta}^{(e)}/2\right) - v_0 \operatorname{sinc}^2\left(\overline{\theta}^{(a)}/2\right)\right) \qquad (21)$$

This result is fully consistent with the previously derived expressions for spontaneous and stimulated emission of FEL and other free electron radiation schemes in the infinite electron quantum wavefunction limit [15].

### Stimulated Emission – coherent photon state

A coherent state represents the classical multi-photon coherent radiation field of a laser beam. In this case the photon state coefficient presentation in terms of Fock states is [31]:

$$\left|\sqrt{v_0}\right\rangle = e^{-v_0/2} \sum_{v=0}^{\infty} \frac{(v_0)^{v/2}}{\sqrt{v!}} |v\rangle \tag{22}$$

And here $v_0$ is the expectation value of the photon number distribution:

$$\sum_v v \left|c_v^{(0)}\right|^2 = v_0 \tag{23}$$

In this case, contrary to the Fock case, substitution of (16a) into (15a) includes a nonvanishing sum of terms $\sum_v \sqrt{v+1}\left(c_{v+1}^{(0)*} c_v^{(0)}\right) = \sum_v \sqrt{v}\left(c_{v-1}^{(0)*} c_v^{(0)}\right) = \sqrt{v_0} \neq 0$.

Therefore, the first order wavepacket-dependent stimulated photon emission of a coherent Glauber state $\Delta v_q^{(1)}$ is nonzero. This is consistent with the conclusion of our earlier semiclassical analysis of this problem [27], and fully expected, since the coherent state represents a classical radiation field. The substitution of (15) into (16) results in for this case for the first order (wavepacket-dependent) and second order (wavepacket-independent) stimulated emission photon contributions:

$$\begin{aligned}\Delta v_q^{(1)} &= \sum_p \left(\sqrt{v_0+1}\,\rho_p^{(1)(e)} + \sqrt{v_0}\,\rho_p^{(1)(a)}\right) \\ \Delta v_q^{(2)} &= \sum_p \left((v_0+1)\,\rho_p^{(2)(e)} - v_0\,\rho_p^{(2)(a)}\right)\end{aligned} \tag{24}$$

where

$$\begin{aligned}\rho_p^{(1)(e,a)} &= 2\Upsilon\left(\frac{p \pm (p_{rec}^{(e,a)} \mp \hbar q_z/2)}{p_0}\right)\text{sinc}\left(\overline{\theta}^{(e,a)}/2\right)\cos\left(\overline{\theta}^{(e,a)}/2 + \phi_0\right)\left(c_p^{(0)*} c_{p \pm p_{rec}^{(e,a)}}^{(0)}\right) \\ \rho_p^{(2)(e,a)} &= \Upsilon^2 \left(\frac{p \pm (p_{rec}^{(e,a)} \mp \hbar q_z/2)}{p_0}\right)^2 \left|c_{p \pm p_{rec}^{(e,a)}}^{(0)}\right|^2\end{aligned} \tag{25}$$

Noted that the second order expression for the stimulated emission is the essentially the same for the coherent state and Fock state, and is given, in the limit of an infinite quantum wavefunction, by the same "FEL gain" [15] wavepacket-independent expression (15). They differ only concerning the first order contribution that is null for single Fock state and finite for a coherent state.

**Discussions**

We now apply the formulation to two specifics examples of quantum electron wavepackets: a single finite size electron wavepacket represented by a Gaussian envelope function, and an optically modulated Gaussian envelope wavepacket [22]

Gaussian electron wavepacket

We consider stimulated emission with a coherent state $\left|\sqrt{v_0}\right\rangle$, interacting with an electron wavepacket of Gaussian distribution, chirped after drift length $L_D$:

$$c_p^{(0)} = \left(2\pi\sigma_{p_0}^2\right)^{-1/4} \exp\left(-\frac{(p-p_0)^2}{4\tilde{\sigma}_p^2(t_D)}\right) e^{i(p_0 L_D - \varepsilon_0 t_D)/\hbar} \tag{26}$$

where $\tilde{\sigma}_p^2(t_D) = \sigma_{p_0}^2\left(1+i\xi t_D\right)^{-1}, \xi = 2\sigma_{p_0}^2/m^*\hbar, L_D = v_0 t_D$. We perform the momentum integration in Eq. 24 for this case, using (25, 26), under the approximation $p_{rec}^{(e,a)}, \hbar q_z, \sigma_{p_0} \ll p_0$ (see Append. A and Ref. [27]). This results in:

$$\Delta v_q^{(1)} = 2\Upsilon\sqrt{v_0} e^{-\Gamma^2/2} \left\{\operatorname{sinc}\left(\overline{\theta}^{(e)}/2\right)\cos\left(\overline{\theta}^{(e)}/2+\phi_0\right) + \operatorname{sinc}\left(\overline{\theta}^{(a)}/2\right)\cos\left(\overline{\theta}^{(a)}/2+\phi_0\right)\right\}$$

$$\Delta v_q^{(2)} = \Upsilon^2\left\{(v_0+1)\operatorname{sinc}^2\left(\overline{\theta}^{(e)}/2\right) - v_0\operatorname{sinc}^2\left(\overline{\theta}^{(a)}/2\right)\right\}$$

$$\tag{27}$$

where we defined the extinction parameter:

$$\Gamma = \left(\frac{\omega}{v_0}\right)\sigma_z(t_D) = \left(\frac{\hbar\omega}{v_0}\right)\frac{\sqrt{1+\xi^2 t_D^2}}{2\sigma_{p_0}} = \Gamma_0\sqrt{1+\xi^2 t_D^2} \tag{28}$$

with $\Gamma_0 = \frac{2\pi}{\beta}\left(\frac{\sigma_{z_0}}{\lambda}\right)$.

This expression is the key result of this publication. While phase-independent expressions for stimulated emission like $(\Delta v_q^{(2)})$ (eq. 27b) can be found in the early literature [15,17-19], the first order phase-dependent term (27a) is new.

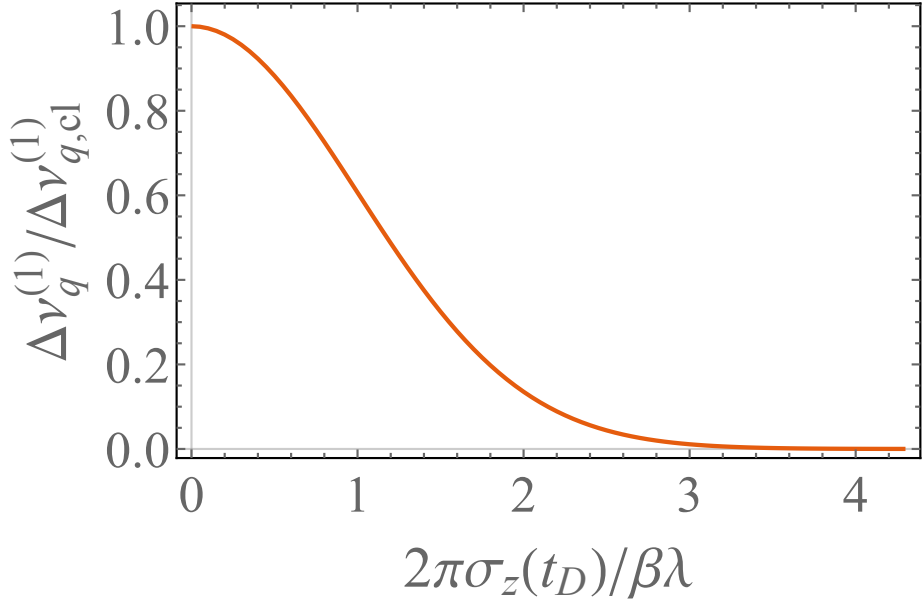

*Fig.3: The wavepacket-dependent photon emission rate as a function of wavepacket size for unmodulated electron wavepacket interacting with a coherent laser beam.*

In the limit of negligible interaction recoil $\varepsilon = \delta\left(\frac{\omega}{v_0}\right)L \ll 1$, $\bar{\theta}^{(e)} = \bar{\theta}^{(a)} = \bar{\theta}$, Eq. 27 reduces to:

$$\Delta v_q^{(1)} = 4\Upsilon\sqrt{v_0}e^{-\Gamma^2/2}\mathrm{sinc}\left(\bar{\theta}/2\right)\cos\left(\bar{\theta}/2+\phi_0\right) \quad (28A)$$

with $\Upsilon = e\mathcal{E}_{qz0}L/4\hbar\omega$. Substituting $\sqrt{v_0}\mathcal{E}_{qz0} = E_{z0} = E_{cl}$, which is the classical axial slow wave field component of the incident radiation wave, we get that the phase-dependent radiation increment is:

$$\hbar\omega\Delta v_q^{(1)} = (eE_{cl}L)e^{-\Gamma^2/2}\mathrm{sinc}\left(\frac{\bar{\theta}}{2}\right)\cos\left(\frac{\bar{\theta}}{2}+\phi_0\right), \quad (29)$$

This result restores the semi-classical expression for electron wavepacket acceleration/deceleration that was derived from solution of Schrodinger equation for the electron [27], and confirms the electron-wave energy conservation spectral reciprocity relation $\Delta v_q^{(1)} + \Delta W_e^{acc}/\hbar\omega = 0$ [28]. The functional dependence of the phase-dependent photon emission (29) on the wavepacket size is shown in Fig. 3 as a function of $\Gamma = 2\pi\sigma_z(t_D)/\beta_0\lambda$ (28). It is suggested that measurement of this dependence can be used for evaluating the wavepacket size $\sigma_z(t_D)$ at the entrance to the interaction region.

Comparison of the first-order wavepacket-dependent stimulated emission expression (28A) that is proportional to $\Upsilon$ ($\mathcal{E}_{qz0}$) and the spontaneous emission expression (18) that is proportional to $\Upsilon^2$ ($\mathcal{E}_{qz0}^2$) suggests a form of "Einstein relation" [28]:

$$\frac{\left(\Delta v_q^{(1)}\right)^2}{\Delta v_{q,SP}} = 16 v_0 e^{-\Gamma^2} \cos^2\left(\frac{\bar{\theta}}{2} + \phi_0\right) \qquad (29A)$$

This is a universal relation that is independent of the interaction structure, because the normalized mode component $\mathcal{E}_{qz0}$ cancels out. It is useful for estimating the wavepacket-dependent ($\Gamma \sim 1$) and the point particle ($\Gamma \ll 1$) stimulated emissions based on measurement of the spontaneous emission.

Comparison of (28A) to (18) also reveals that while the wavepacket-dependent stimulated emission vanishes in the range $\Gamma(L_D) \gg 1$ (and absolutely so at any drift distance from its source $L_D > z_G = \frac{\beta_0^3 \gamma_0^3}{\pi} \frac{\lambda^2}{\lambda_c}$ [27]), the quantum spontaneous emission always exists, independently of $\Gamma$. Therefore, observation of classical single point-particle emission and recognizing the transition of wavepacket–dependent stimulated emission from the classical to the quantum limit in the regime $\Gamma \sim 1$, require overcoming a signal to noise ratio condition S/N>1, where

$$\frac{S}{N} \equiv \left(\frac{\Delta v_q^{(1)}}{\Delta v_{q,SP}}\right)\Bigg|_{max} = \frac{4v_0}{\Upsilon} \qquad (29B)$$

Note that $\Upsilon = e\mathcal{E}_{qz0}L/4\hbar\omega$ is structure-dependent, and has to be evaluated from the mode quantization condition (2A, 2B).

## Modulated Gaussian electron wavepacket

Now we consider the case where the initial state is an optically modulated Gaussian electron wavepacket. Such an electron wavefunction can be generated by multiphoton emission/absorption from a laser beam of frequency $\omega_b$. After a drift length $L_D$ in dispersive free space, its multi-harmonic momentum distribution is chirped [22]:

$$c_p^{(0)} = \left(2\pi\sigma_{p_0}^2\right)^{-1/4} \sum_{n=-\infty}^{\infty} J_n(2|g|) \exp\left(-\frac{(p-p_0-n\delta_p)^2}{4\sigma_{p_0}^2}\right) e^{-i(p-p_0)^2 t_D/2m^*\hbar} e^{i(p_0 L_D - \varepsilon_0 t_D)/\hbar} \quad (30)$$

where $\delta_p = \hbar\omega_b/v_0$ is the multi-photon emission/absorption electron momentum-recoil quantum at the modulation point. The detailed derivation of (29) can be found in [22,28], where it is shown that (30) represents an expectation-value-density modulated wvepacket. Substituting this expression into eq. 25, and performing the integration over p in eq. 24 (see Append. B), one obtains:

$$\Delta\nu_q^{(1)} = 2\Upsilon\sqrt{\nu_0}\left\{B^{(e)}\operatorname{sinc}\left(\bar\theta^{(e)}/2\right)\cos\left(\bar\theta^{(e)}/2+\phi_0\right) + B^{(a)}\operatorname{sinc}\left(\bar\theta^{(a)}/2\right)\cos\left(\bar\theta^{(a)}/2+\phi_0\right)\right\}$$
$$\Delta\nu_q^{(2)} = \Upsilon^2\left\{(\nu_0+1)\operatorname{sinc}^2\left(\bar\theta^{(e)}/2\right) - \nu_0\operatorname{sinc}^2\left(\bar\theta^{(a)}/2\right)\right\} \quad (31)$$

The second order term is the same as in the case of an unmodulated electron wavepacket (27), where we used the following mathematical sum-rule:

$$\sum_{n,m=-\infty}^{\infty} J_n(2|g|) J_m(2|g|) \exp\left(-\frac{(n-m)^2 \delta_p^2}{8\sigma_{p_0}^2}\right) = 1$$

. This term is therefore the conventional plane wave expression for quantum spontaneous and stimulated emission [15], independent of the wavepacket dimensions and internal distribution. On the other hand, the first order term is dependent on both the wavepacket dimension and modulation parameters through the bunching decay parameters:

$$B^{(e,a)} = \exp\left(-\frac{(1+\xi^2 t_D^2)p_{rec}^{(e,a)2}}{8\sigma_{p_0}^2}\right) \sum_{n,m=-\infty}^{\infty} J_n(2|g|) J_m(2|g|) \exp\left(-\frac{(n-m)^2\delta_p^2}{8\sigma_{p_0}^2} + \frac{(n-m)\delta_p p_{rec}^{(e,a)}}{4\sigma_{p_0}^2} \pm \frac{i(n+m)\delta_p p_{rec}^{(e,a)} t_D}{2m^*\hbar}\right)$$

(32)

where we approximate $p_{rec}^{(e,a)} = p_{rec}^{(0)} = \hbar\omega/v_0$. As in the case of unmodulated wavepacket (29), we express (31a) in the limit of negligible interaction recoil $\varepsilon = \delta\left(\frac{\omega}{v_0}\right) L \ll 1$,

$\bar{\theta}^{(e)} = \bar{\theta}^{(a)} = \bar{\theta}$:

$$\Delta v_{q,\text{mod}}^{(1)} = \left(\frac{eE_{cl}L}{\hbar\omega}\right) B(\omega) \text{sinc}\left(\frac{\bar{\theta}}{2}\right) \cos\left(\frac{\bar{\theta}}{2} + \phi_0\right) \quad (33)$$

where (see Append. B):

$$B(\omega) = \frac{B^{(e)} + B^{(a)}}{2} = \sum_{l=-\infty}^{\infty} B_l \exp\left(-\frac{(\omega - l\omega_b)^2 \sigma_\tau^2(t_D)}{2}\right) \quad (34)$$

Note that the expression has an index symmetry ($n \to -n, m \to -m$) and with the relations of Bessel functions $J_{-n}(2|g|)J_{-m}(2|g|) = (-1)^{n+m} J_n(2|g|)J_m(2|g|)$, one obtains only the terms when $n \pm m$ is even have contribute to the summation. The $l^{\text{th}}$-order bunching parameter when $p_{rec}^{(0)} = (n-m)\delta_p \Rightarrow \omega = l\omega_b$ with $l = n - m$ is then given by

$$B_l \approx \sum_{n=-\infty}^{\infty} J_n(2|g|) J_{n-l}(2|g|) \exp\left(-\frac{l^2(\delta_p \xi t_D)^2}{8\sigma_{p_0}^2}\right) \cos\left(\frac{(2n-l)l\delta_p^2 t_D}{2m^*\hbar}\right) \quad (35)$$

and $\delta_p = \hbar\omega_b/v_0$.

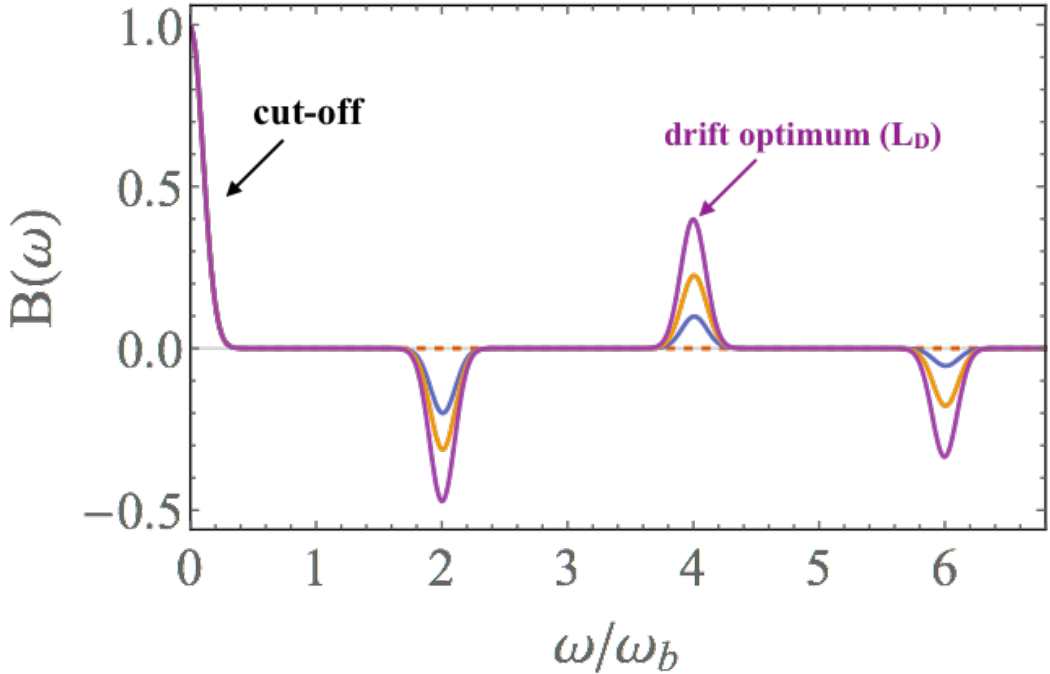

Fig.4: The bunching parameter with (blue curve)/without (red curve) the optimal drift length ($L_D$) of stimulated radiance and its dependence beyond the spectral cut-off for

*modulated electron wavepackets. The parameters are* $\omega_b \sigma_t(L_D) = 4$.

The multi-harmonic spectrum of the decay parameter $B(\omega)$ is shown in Fig. 4 for a case $\omega \sigma_t(t_D) = 4 \gg 1$ that corresponds to a long wavepacket with internal modulation and separable harmonics. The harmonic amplitudes were calculated for the drift length for which the modulation amplitude is maximal [22,28]. Note that the symmetry of eq. 32 is such that only even harmonics of the modulation frequency appear in the spectrum. Also, notice that (34) reduces to (29) in the limit of no modulation (g=0). When g>0, the fundamental harmonic decays with decay constant $\Gamma(t_D)$ as in (27,28). However, the higher harmonic spectral spots l>1 show emission/absorption beyond the cutoff frequency condition $\Gamma > 1$. Comparative measurement of the emission spectrum with modulation (Fig. 4) and without (Fig. 3) may be helpful in the measurement of the wavepacket size.

## Conclusions

The main results of the present analysis are summarized in Table 1 for both cases of finite size unmodulated and modulated quantum electron wavepackets. Solving for the interaction of an electron wavepacket with quantized radiation, we identified two additive contributions to the photon emission: wavepacket-dependent ($\Delta v_q^{(1)}$) and wavepacket independent $\Delta v_q^{(2)}$. The second order term is consistent with the conventional quantum theory for spontaneous and spontaneous emission of free electrons in the infinite quantum wavefunction regime [15]. The first order term $\Delta v_q^{(1)}$ is innovative. It predicts wavepacket-dependence of stimulated emission when the interacting radiation state is coherent (Glauber) state, consistently and complementarily with earlier predictions of electron wavepacket-dependent acceleration/deceleration, based on semi-classical (electron Schrodinger equation) analysis [27]. A Fock state interacting wave has null contribution to the first order term $\Delta v_q^{(1)}$. This includes also the case of vacuum state $v_{q0} = 0$, indicating wavepacket-independence of spontaneous emission in all regimes independently whether the wavepacket is modulated or not.

The main result of this work is the affirmation that the first order stimulated emission spectrum of a wavepacket depends on its size at the entrance to the interaction region $\sigma_z(t_D)$, following an exponential decay scaling (eq. 29, Fig. 3) with a short wavelength cutoff when $\sigma_z(t_D) \sim \beta_0 \lambda$, corresponding to the transition from point-particle classical interaction limit to the quantum electron wavefunction limit. A more intricate stimulated emission spectrum takes place when the wavepacket is optically modulated, displaying wavepacket-dependent characteristics harmonic frequencies spectrum (eq. 34, Fig. 4).

We assert that measurement of the characteristic stimulated emission spectra stimulated emission spectra of modulated and unmodulated quantum electron wavepacket, provides a way for evaluating its size and its internal features. Such measurement can be done by changing the interaction wavelength $\lambda$ or the drift length $L_D = v_0 t_D$ in the range $\sigma_z(t_D) = \sigma_{z0}\sqrt{1+\xi^2 t_D^2}$ which is attainable at short enough drift lengths away from the source $L_D < z_G = \frac{\beta_0^3 \gamma_0^3}{\pi} \frac{\lambda^2}{\lambda_c}$ [27]). We stress, however, that only for simplicity we assumed that the wavepacket size at the entrance to the interaction region is determined by the drift time ($\sigma_z(t_D)$). It has been recently shown that the quantum electron wavepacket phase, size and chirp characteristics are controllable by electron dispersion and optical streaking techniques [33]. Hence, the more general conclusion of this paper is that the stimulated interaction of a free electron wavepacket can be dependent on the history of the electron transport prior to the interaction.

Practical measurement of photon emission and electron energy spectra of single electron radiative interaction events, is a challenging experiment. It may require accumulating multi-interaction events data with wavepacket preselection of the Aharonov-Weidman kind [34] and wavepacket shape formation schemes as in [34]. It certainly also requires satisfaction of a signal-to-noise condition S/N>1, considering the ever present wavepacket-independent noise due to spontaneous emission (29B).

| PHOTON EMISSIONS $\Delta\nu_q = \Delta\nu_q^{(1)} + \Delta\nu_q^{(2)}$ | | Gaussian Wavepacket | Modulated Wavepacket |
|---|---|---|---|
| Spontaneous Emission | Vacuum State | $\Delta\nu_q^{(1)} = 0$ | |
| | | $\Delta\nu_q^{(2)} = \Upsilon^2 \mathrm{sinc}^2(\bar{\theta}^{(e)}/2)$ | |
| Stimulated Emission | Fock State | $\Delta\nu_q^{(1)} = 0$ | $\Delta\nu_q^{(1)} = 0$ |
| | | $\Delta\nu_q^{(2)} = \Upsilon^2 \left\{ (\nu_0 + 1)\mathrm{sinc}^2(\bar{\theta}^{(e)}/2) - \nu_0 \mathrm{sinc}^2(\bar{\theta}^{(a)}/2) \right\}$ | |
| | Coherent State | $\Delta\nu_q^{(1)} = \left(\frac{eE_{cl}L}{\hbar\omega}\right) e^{-\Gamma^2/2} \mathrm{sinc}(\bar{\theta}/2) \cos(\bar{\theta}/2 + \phi_0)$ | $\Delta\nu_q^{(1)} = \left(\frac{eE_{cl}L}{\hbar\omega}\right) B(\omega) \mathrm{sinc}(\bar{\theta}/2) \cos(\bar{\theta}/2 + \phi_0)$ |
| | | $\Delta\nu_q^{(2)} = \Upsilon^2 \left\{ (\nu_0 + 1)\mathrm{sinc}^2(\bar{\theta}^{(e)}/2) - \nu_0 \mathrm{sinc}^2(\bar{\theta}^{(a)}/2) \right\}$ | |

*Table 1: A gallery of phase-dependent and phase-independent photon emission rates of unmodulated and modulated quantum electron wavepackets.*

## Acknowlegements

We acknowledge W. Scheleich and P. Kling for useful discussions and comments. The work was supported in parts by DIP (German-Israeli Project Cooperation) and by the PBC program of the Israel council of higher education. Correspondance and requests for materials should be addressed to A. G.(gover@eng.tau.ac.il).

# Supplementary Materials:

1. **Derivation of photon emission in the case of a Gaussian wavepacket**

To derive the photon emission expression (28), the integration over p in eq. 24 should be carried out with the Gaussian distribution function of the drifted electron amplitude in momentum space (26). For the phase-independent second order photon emission $\Delta v_q^{(2)}$, this involves the following integrations:

$$\sum_p \left\{ \left( \frac{p + p_{rec}^{(e)} - \hbar q_z/2}{p_0} \right)^2 \left| c_{p+p_{rec}^{(e)}}^{(0)} \right|^2 \right\}$$

$$= \left( 2\pi\sigma_{p_0}^2 \right)^{-1/2} \int dp \left( \frac{p + p_{rec}^{(e)} - \hbar q_z/2}{p_0} \right)^2 \exp\left( -\frac{(p + p_{rec}^{(e)} - p_0)^2}{2\sigma_{p_0}^2} \right)$$

$$= \left( 1 - \frac{\hbar q_z}{2 p_0} \right)^2 + \left( \frac{\sigma_{p_0}}{p_0} \right)^2 \approx 1$$

and similarly, for the absorption term:

$$\sum_p \left\{ \left( \frac{p - (p_{rec}^{(a)} - \hbar q_z/2)}{p_0} \right)^2 \left| c_{p-p_{rec}^{(a)}}^{(0)} \right|^2 \right\} = \left( 1 + \frac{\hbar q_z}{2 p_0} \right)^2 + \left( \frac{\sigma_{p_0}}{p_0} \right)^2 \approx 1$$

For the phase-dependent first-order photon emission part ($\Delta v_q^{(1)}$):

$$\sum_p \left\{ \left( \frac{p + (p_{rec}^{(e)} - \hbar q_z/2)}{p_0} \right) \left( c_p^{(0)*} c_{p+p_{rec}^{(e)}}^{(0)} \right) \right\}$$

$$= \left( 2\pi\sigma_{p_0}^2 \right)^{-1/2} \int dp \left( \frac{p + p_{rec}^{(e)} - \hbar q_z/2}{p_0} \right) \exp\left( -\frac{(p - p_0)^2}{4\sigma_{p_0}^2 (1 - i\xi t_D)^{-1}} \right) \exp\left( -\frac{(p + p_{rec}^{(e)} - p_0)^2}{4\sigma_{p_0}^2 (1 + i\xi t_D)^{-1}} \right)$$

$$= e^{-\Gamma^2/2} \left( 1 + \frac{p_{rec}^{(e)} - \hbar q_z + i p_{rec}^{(e)} \xi t_D}{2 p_0} \right) \approx e^{-\Gamma^2/2}$$

and similarly, for the absorption term:

$$\sum_p \left\{ \left( \frac{p - (p_{rec}^{(a)} - \hbar q_z/2)}{p_0} \right) \left( c_p^{(0)*} c_{p-p_{rec}^{(a)}}^{(0)} \right) \right\} = e^{-\Gamma^2/2} \left( 1 - \frac{p_{rec}^{(e)} - \hbar q_z + i p_{rec}^{(e)} \xi t_D}{2 p_0} \right) \approx e^{-\Gamma^2/2}$$

where we define the decay parameter

$$\Gamma = \left(\frac{\omega}{v_0}\right)\sigma_z(t_D) = \left(\frac{\hbar\omega}{v_0}\right)\frac{\sqrt{1+\xi^2 t_D^2}}{2\sigma_{p_0}} = \Gamma_0\sqrt{1+\xi^2 t_D^2} \quad \text{and} \quad \Gamma_0 = \frac{2\pi}{\beta_0}\left(\frac{\sigma_z}{\lambda}\right), \quad \xi = \frac{2\sigma_{p_0}^2}{\gamma_0^3 m\hbar}$$

(eq. 28).

Note that in all cases we took the approximation $p_{rec}^{(e,a)}, \hbar q_z, \sigma_{p_0} \ll p_0$ in the last steps of calculation. Also, note that the imaginary part may contribute to an additional phase to the cosine function in the case of very long drift time $t_D$.

## 2. Derivation of photon emission in the case of a modulated Gaussian wavepacket

To derive the photon emission expression (31), the integration over p in eqs. 24,25 should be carried out the integration over p in eq. 24 should be carried out with the modulated Gaussian distribution function of the drifted electron amplitude in momentum space (30). For the phase-independent second order photon emission $\Delta v_q^{(2)}$, this involves the following integrations:

$$\sum_p \left\{ \left(\frac{p + p_{rec}^{(e)} - \hbar q_z/2}{p_0}\right)^2 \left|c_{p+p_{rec}^{(e)}}^{(0)}\right|^2 \right\}$$

$$= \left(2\pi\sigma_{p_0}^2\right)^{-1/2} \sum_{n,m=-\infty}^{\infty} J_n(2|g|) J_m(2|g|) \int dp \left(\frac{p+p_{rec}^{(e)}-\hbar q_z/2}{p_0}\right)^2 \exp\left(-\frac{(p+p_{rec}^{(e)}-p_0-n\delta_p)^2}{4\sigma_{p_0}^2}\right) \exp\left(-\frac{(p+p_{rec}^{(e)}-p_0-m\delta_p)^2}{4\sigma_{p_0}^2}\right)$$

$$= \left(\left(1-\frac{(n+m)\delta_p + \hbar q_z}{2p_0}\right)^2 + \left(\frac{\sigma_{p_0}}{p_0}\right)^2\right) \sum_{n,m=-\infty}^{\infty} J_n(2|g|) J_m(2|g|) \exp\left(-\frac{(n-m)^2\delta_p^2}{8\sigma_{p_0}^2}\right)$$

$$\approx \sum_{n,m=-\infty}^{\infty} J_n(2|g|) J_m(2|g|) \exp\left(-\frac{(n-m)^2\delta_p^2}{8\sigma_{p_0}^2}\right) = 1$$

and similarly:

$$\sum_p \left\{ \left(\frac{p-(p_{rec}^{(a)} - \hbar q_z/2)}{p_0}\right)^2 \left|c_{p-p_{rec}^{(a)}}^{(0)}\right|^2 \right\} \approx \sum_{n,m=-\infty}^{\infty} J_n(2|g|) J_m(2|g|) \exp\left(-\frac{(n-m)^2\delta_p^2}{8\sigma_{p_0}^2}\right) = 1$$

Here we took the approximation $p_{rec}^{(e,a)}, |n+m|\delta_p = |n+m|\hbar\omega_b/v_0, \hbar q_z, \sigma_{p_0} \ll p_0$ in the last steps of calculation and adapted an identity relation of Bessel functions.

Using the same approximations for the phase-dependent first-order photon emission term ($\Delta \nu_q^{(1)}$):

$$\sum_p \left\{ \left( \frac{p + (p_{rec}^{(e)} - \hbar q_z/2)}{p_0} \right) \left( c_p^{(0)*} c_{p+p_{rec}^{(e)}}^{(0)} \right) \right\}$$

$$= (2\pi\sigma_{p_0}^2)^{-1/2} \sum_{n,m=-\infty}^{\infty} J_n(2|g|) J_m(2|g|) \int dp \left( \frac{p + p_{rec}^{(e)} - \hbar q_z/2}{p_0} \right) \exp\left( -\frac{(p - p_0 - n\delta_p)^2}{4\sigma_{p_0}^2} - \frac{(p + p_{rec}^{(e)} - p_0 - m\delta_p)^2}{4\sigma_{p_0}^2} \right) e^{\frac{i(p - p_0 - n\delta_p)^2 t_D}{2m^*\hbar} - \frac{i(p + p_{rec}^{(e)} - p_0 - m\delta_p)^2 t_D}{2m^*\hbar}}$$

$$= \left( 1 + \frac{(n+m)\delta_p + p_{rec}^{(e)} - \hbar q_z + i p_{rec}^{(e)} \xi t_D}{2 p_0} \right) e^{-\frac{(1 + \xi^2 t_D^2) p_{rec}^{(e)2}}{8\sigma_{p_0}^2}} \sum_{n,m=-\infty}^{\infty} J_n(2|g|) J_m(2|g|) \exp\left( -\frac{(n-m)^2 \delta_p^2}{8\sigma_{p_0}^2} + \frac{(n-m)\delta_p p_{rec}^{(e)}}{4\sigma_{p_0}^2} \pm \frac{i(n+m)\delta_p p_{rec}^{(e)} t_D}{2m^*\hbar} \right)$$

$$\approx e^{-\frac{(1 + \xi^2 t_D^2) p_{rec}^{(e)2}}{8\sigma_{p_0}^2}} \sum_{n,m=-\infty}^{\infty} J_n(2|g|) J_m(2|g|) \exp\left( -\frac{(n-m)^2 \delta_p^2}{8\sigma_{p_0}^2} + \frac{(n-m)\delta_p p_{rec}^{(e)}}{4\sigma_{p_0}^2} + \frac{i(n+m)\delta_p p_{rec}^{(e)} t_D}{2m^*\hbar} \right)$$

and similarly, for the absorption term:

$$\sum_p \left\{ \left( \frac{p - (p_{rec}^{(a)} - \hbar q_z/2)}{p_0} \right) \left( c_p^{(0)*} c_{p - p_{rec}^{(a)}}^{(0)} \right) \right\}$$

$$\approx e^{-\frac{(1 + \xi^2 t_D^2) p_{rec}^{(a)2}}{8\sigma_{p_0}^2}} \sum_{n,m=-\infty}^{\infty} J_n(2|g|) J_m(2|g|) \exp\left( -\frac{(n-m)^2 \delta_p^2}{8\sigma_{p_0}^2} - \frac{(n-m)\delta_p p_{rec}^{(a)}}{4\sigma_{p_0}^2} - \frac{i(n+m)\delta_p p_{rec}^{(e)} t_D}{2m^*\hbar} \right)$$

We note that the expression for the second order photon emission $\Delta \nu_q^{(2)}$ (eq. 31b), including the expression for spontaneous emission ($\nu_0 = 0$), is identical to the expression for the unmodulated wavepacket (27), namely – the modulation, as well as the wavepacket dimension do not affect the spontaneous emission spectrum at all. Also, note that the expression for the first order photon emission $\Delta \nu_q^{(1)}$ (eq. 31a) reduces to the expression of the wavepacket-dependent first order term of the unmodulated wavepacket (eq. 29) in the limit of diminished modulation parameter $2|g| \to 0$, where the identity $e^{-(1 + \xi^2 t_D^2) p_{rec}^{(e)2} / 8\sigma_{p_0}^2} = e^{-\Gamma^2/2}$ recovers the the spectral cut-off factor in eq. 29.

To derive the first order wavepacket size and modulation-dependent photon emission expression (33,34), we substitute in eq. 32:

$$B^{(e,a)} = \exp\left( -\frac{(1 + \xi^2 t_D^2) p_{rec}^{(e,a)2}}{8\sigma_{p_0}^2} \right) \sum_{n,m=-\infty}^{\infty} J_n(2|g|) J_m(2|g|) \exp\left( -\frac{(n-m)^2 \delta_p^2}{8\sigma_{p_0}^2} + \frac{(n-m)\delta_p p_{rec}^{(e,a)}}{4\sigma_{p_0}^2} \pm \frac{i(n+m)\delta_p p_{rec}^{(e,a)} t_D}{2m^*\hbar} \right)$$

the approximation $p_{rec}^{(e,a)} = p_{rec}^{(0)} = \hbar\omega / v_0$, resulting in:

$$B(\omega) = \frac{B^{(e)} + B^{(a)}}{2} = e^{-\Gamma^2/2} \sum_{n,m=-\infty}^{\infty} J_n(2|g|) J_m(2|g|) \exp\left(-\frac{(n-m)^2 \delta_p^2}{8\sigma_{p0}^2} + \frac{(n-m)\delta_p p_{rec}^{(0)}}{4\sigma_{p0}^2}\right) \cos\left(\frac{(n+m)\delta_p p_{rec}^{(e,a)} t_D}{2m^*\hbar}\right)$$

$$= \sum_{l=-\infty}^{\infty} \left( \sum_{n=-\infty}^{\infty} J_n(2|g|) J_{n-l}(2|g|) \exp\left(-\frac{\left(p_{rec}^{(0)} - l\delta_p + l\delta_p\right)^2 \xi^2 t_D^2}{8\sigma_{p0}^2}\right) \cos\left(\frac{(2n-l)\delta_p p_{rec}^{(e,a)} t_D}{2m^*\hbar}\right) \right) \exp\left(-\frac{\left(l\delta_p - p_{rec}^{(0)}\right)^2}{8\sigma_{p0}^2}\right)$$

$$= \sum_{l=-\infty}^{\infty} B_l \exp\left(-\frac{\left(p_{rec}^{(0)} - l\delta_p\right)^2}{8\sigma_{p0}^2 \left(1+\xi^2 t_D^2\right)^{-1}}\right) = \sum_{l=-\infty}^{\infty} B_l \exp\left(-\frac{(\omega - l\omega_b)^2 \sigma_t^2(t_D)}{2}\right)$$

With $p_{rec}^{(0)} = (n-m)\delta_p \Rightarrow \omega = l\omega_b$ and $l = n-m$ being the micro-bunching harmonic order. The $l^{th}$-order bunching parameter is given by:

$$B_l = \sum_{n=-\infty}^{\infty} J_n(2|g|) J_{n-l}(2|g|) \exp\left(\frac{-l\delta_p\left(2p_{rec}^{(0)} - l\delta_p\right)\xi^2 t_D^2}{8\sigma_{p0}^2}\right) \cos\left(\frac{(2n-l)\delta_p p_{rec}^{(e,a)} t_D}{2m^*\hbar}\right)$$

$$\approx \sum_{n=-\infty}^{\infty} J_n(2|g|) J_{n-l}(2|g|) \exp\left(-\frac{l^2 (\delta_p \xi t_D)^2}{8\sigma_{p0}^2}\right) \cos\left(\frac{(2n-l)l\delta_p^2 t_D}{2m^*\hbar}\right)$$

where $\delta_p = \hbar\omega_b/v_0$. This frequency dependence of the bunching parameter factor in the first order stimulated emission explains the remarkable resonant radiative "spots" at $\omega = l\omega_b$ in the stimulated emission spectrum (Fig.4) beyond the frequency cut-off of $l = 0$, reflecting the interior micro-structure of electron wavepacket.